\documentclass{ws-ijtaf}

\usepackage{footnote}
\usepackage{rotating}
\usepackage{hhline}
\usepackage[normalem]{ulem}
\usepackage{graphicx}
\usepackage{chngcntr}
\usepackage{color}

\begin{document}

\markboth{Grasselli and Lipton}
{The Broad Consequences of Narrow Banking}

\catchline{}{}{}{}{}

\title{THE BROAD CONSEQUENCES OF NARROW BANKING}

\author{MATHEUS R. GRASSELLI}
\address{Department of Mathematics and Statistics, McMaster University \\
1280 Main Street West, Hamilton ON L8S 4K1, Canada \\
\email{grasselli@math.mcmaster.ca}}

\author{ALEXANDER LIPTON}
\address{Connection Science, Massachusetts Institute of Technology \\
 77 Massachusetts Ave, Cambridge, MA 02139, USA \\
\email{alexlipt@mit.edu}}

\maketitle


\begin{abstract}
We investigate the macroeconomic consequences of narrow banking in the
context of stock-flow consistent models. We begin with an extension of the Goodwin-Keen model incorporating 
time deposits, government bills, cash, and central bank reserves to the base model with loans and demand deposits and 
use it to describe a fractional reserve banking system. We then
characterize narrow banking by a full reserve requirement on demand deposits and describe the resulting 
separation between the payment system and lending functions of the resulting banking sector. By way of numerical examples, we 
explore the properties of fractional and full reserve versions of the model and compare their asymptotic properties. We find that narrow banking 
does not lead to any loss in economic growth when the models converge to a finite equilibrium, while allowing for more direct monitoring and prevention of
financial breakdowns in the case of explosive asymptotic behaviour.   
\end{abstract}

\keywords{narrow banking; macroeconomic dynamics; debt-financed investment; financial stability.}

\section{Introduction}

\label{introduction}

Narrow banking is a recurrent theme in economics, especially after periods
of financial turbulence. For example, in the wake of the Great Depression,
several prominent economists proposed a set of reforms known as the Chicago
Plan, which included the requirement that banks hold reserves matching the
amount of demand deposits (see \cite{Phillips1996}). The Banking Act of 1935
adopted different measures to promote stability of the banking sector, such
as deposit insurance and the separation of commercial and investment
banking, but the idea of narrow banking never went away -- and neither did
financial crises. After the Global Financial Crisis of 2008 the idea came to
the fore again, with outlets as diverse as the International Monetary Fund and the Positive Money movement 
re-examining the benefits of narrow banking for the current financial system (see \cite{KumhofBenes2012,DysonHodgsonvanLerven2016}), 
and in at least one case a government explicitly dedicating resources to debate the idea (see \cite{KPMG2016}). 

More recently, narrow banking attracted a lot of attention in the context 
the cryptocurrency boom, especially in conjunction with potential
introduction of central bank issued digital currencies \cite{BarrdearKumhof2016, Lipton2016, LiptonPentlandHardjono2018}. It even 
made headlines outside academic circles (see \cite{Levine2018}) when the Federal Reserve Bank of New York denied an application to open an account by 
TNB USA Inc., which stands 
for The Narrow Bank\footnote{In 2016-17, one of the authors also tried to build a narrow bank in practice - see 
Burkov, S., Dembo, R., and Lipton, A., 2017 Account platform for a distributed network of nodes. U.S. Patent Application.}

The main feature of a narrow bank is its assets mix. By definition, these
assets can include only marketable low-risk liquid (government) securities, cash, or central bank reserves 
in the amount exceeding its deposit base. As a result,
such a bank is immune to market, credit and liquidity risks, and can only be
toppled by operational failures that can be minimized, though not eliminated, by using state-of-the-art technology.
Consequently, the liabilities of a narrow bank, in the form of demand deposits, are equivalent to 
currency and provide a maximally safe payment system. Accordingly, a narrow bank does not require deposit insurance with all its complex and poorly
understood effects on the system as a whole, including not so subtle moral hazards. 

Because loans and other risky securities are excluded from the asset mix of a narrow bank, lending has to be performed 
by specially constructed lending facilities, which would need to raise funds from the private sector and the
government \emph{before} lending them out, in contrast to fractional reserve
banks who \emph{simultaneously} create funds and lend them. In other words, narrow banking separates two functions 
that are traditionally performed together in conventional banking: lending and the payment system. 

In practice, a narrow bank and a lending facility as defined above can be combined into a single business, much like the same company can 
sell both computers and cellphones. All that is necessary is that any bank accepting demand deposits as liabilities be required to hold an equal or 
greater amount of central bank reserves as assets. In what follows, we take this full reserve requirement as the operational definition of narrow banking and 
compare it with the current practice of fractional reserve banking. 

We perform the analysis in a stock-flow consistent framework similar to 
that of \cite{Laina2015} but using the model for debt-financed investment proposed in \cite{Keen1995} as a 
starting point. Apart from the usual technical differences between discrete and continuous-time models, the main advantage 
of this model is that it allows us to draw conclusions for a growing economy, whereas \cite{Laina2015} is limited to the zero-growth case. 

Our two main findings are: (1) narrow banking does not impede growth and (2) whereas it does not entirely prevent financial crises either, it allows 
for more direct monitoring and preventive intervention by the government. The first finding it significant because a common objection to narrow banking is that removing 
the money-creation capacity from private banks would lead to a shortage of available funds to finance investment and promote growth. Our results show that this is not the case, with a combination of private and public funds being sufficient to finance investment and lead to economic growth at the same equilibrium rates in both the full and fractional reserve cases. In our view, this result alone is enough to justify a much wider discussion of narrow banking than has occurred so far, as the clear advantages mentioned earlier, such as the reduced need for deposit insurance, do not need to be necessarily weighed against losses in economic growth. 

Our second finding is more subtle. In the context of the model analyzed in this paper, a financial crisis is associated with an equilibrium with exploding ratios of private debt and accompanying ever decreasing employment, wage share, and output. We find that such equilibria are present in both the full and fractional reserve cases and are moreover associated with exploding ratios of government {\em lending} to the private sector. In the narrow banking case, however, this last variable - namely the ratio of 
government lending to GDP - exhibits a clearly explosive behaviour much sooner than in the fractional reserve case. Because this is an indicator that is under direct control of the government (as opposed to capital or leverage ratios in the private sector, for example), it is much easier for regulators in the narrow banking case to detect the onset of a crisis and take measures to prevent it. 

The rest of the paper is organized as follows. In Section \ref{fractional},
we extend to \cite{Keen1995} model by introducing both demand and time deposits as liabilities of the banking sector, as
well as reserves and government bills, in addition to loans, as assets of this sector. Accordingly, we introduce a central bank 
conducting monetary policy in order to achieve a policy rate on government bills and provide the banking sector with the 
required amount of reserves. As in the \cite{Keen1995} model, the key decision variable of the private sector is the amount of 
investment by firms, whereas households adjust their consumption accordingly, with the only added feature of a portfolio selection for 
households along the lines of \cite{Tobin1969}. Finally, we assume a simplified fiscal policy in the form of government spending and taxation as 
constant proportions of output. 

In Section \ref{narrow} we modify the model by imposing 100\% reserve
requirements for banks. This has the effect of limiting bank lending, which
now needs to be entirely financed by equity and other borrowing. In the
present model, a capital adequacy ratio smaller than 1 can only be
maintained by borrowing from the central bank, which can in effect control
the total amount of bank lending.

In Section \ref{numerical} we study the models introduced in Sections \ref{fractional} and \ref{narrow} and show how economy can develop under both
beneficial and adverse circumstances by way of examples illustrating the properties described earlier. In Section \ref{conclusion} we review our conclusions and outline future research directions.

\section{Fractional reserve banking}

\label{fractional}

We consider a five-sector closed economy consisting of firms, banks,
households, a government sector and a central bank as summarized in Table %
\ref{table}. As it is typical in stock-flow consistent models, the balance
sheet, transactions, and flow of funds depicted in this table already
encapsulate a lot of the structure in the model, so we start by describing
each item in some detail.

\begin{table}[h]
\tbl{Balance sheets, transactions and flow of funds matrices.}  {\begin{tabular}{|l|c|cc|c|c|c|c|}
\hline
& Households & \multicolumn{2}{|c|}{Firms} & Banks & Gov & CB & Row sum \\ 
\hline
\textbf{Balance Sheet} &  &  &  &  &  &  &  \\ 
Capital stock &  & \multicolumn{2}{|c|}{$+pK$} &  &  &  & $+pK$ \\ 
Loans &  & \multicolumn{2}{|c|}{$-L$} & $+L$ &  &  & 0 \\ 
Cash & $+H$ &  &  &  &  & $-H$ & 0 \\ 
Treasury bills & $+\Theta_h$ &  &  & $+\Theta_b$ & $-\Theta $ & $+\Theta_{cb}
$ & 0 \\ 
Demand deposits & $+M_h$ & \multicolumn{2}{|c|}{$+M_f$} & $-M$ &  &  & 0 \\ 
Tme deposits & $+D$ &  &  & $-D$ &  &  & 0 \\ 
Reserves &  &  &  & $+R$ &  & $-R$ & 0 \\ \hline
Column sum (net worth) & $X_h$ & \multicolumn{2}{|c|}{$X_f$} & $X_b$ & $X_g$
& $0$ & $pK$ \\ \hline\hline
\textbf{Transactions} &  & current & capital &  &  &  &  \\ 
Consumption & $-pC$ & $+pC$ &  &  &  &  & 0 \\ 
Gov Spending &  & $+pG$ &  &  & $-pG$ &  & 0 \\ 
Capital Investment &  & $+pI$ & $-pI$ &  &  &  & 0 \\ 
Accounting memo [GDP] &  & [$pY$] &  &  &  &  &  \\ 
Wages & $+W$ & $-W$ &  &  &  &  & 0 \\ 
Taxes &  & $-pT$ &  &  & $+pT$ &  & 0 \\ 
Interest on loans &  & $-rL$ &  & $+rL$ &  &  & 0 \\ 
Depreciation &  & $-p\delta K$ & $+p\delta K$ &  &  &  & 0 \\ 
Interest on bills & $+r_\theta\Theta_h$ &  &  & $+r_\theta\Theta_b$ & $-r_\theta\Theta$ & $+r_\theta\Theta_{cb}$ & 0 \\ 
Interest on demand deposits & $+r_mM_h$ & $+r_mM_f$ &  & $-r_mM$ &  &  & 0
\\ 
Interest on time deposits & $+r_dD$ &  &  & $-r_d D$ &  &  & 0 \\ 
Bank dividends & $+\Delta_b$ &  &  & $-\Delta_b$ &  &  & 0 \\ 
CB profits &  &  &  &  & $+\Pi_{cb}$ & $-\Pi_{cb}$ & 0 \\ \hline
Column sum (financial balances) & $S_h$ & $S_f$ & $-p(I-\delta K)$ & $S_b$ & 
$S_g$ & 0 & 0 \\ \hline\hline
\textbf{Flow of Funds} &  &  &  &  &  &  &  \\ 
Change in capital stock &  & \multicolumn{2}{|c|}{$+p(I-\delta K)$} &  &  & 
& $+p(I-\delta K)$ \\ 
Change in loans &  & \multicolumn{2}{|c|}{$-\dot{L}$} & $+\dot{L}$ &  &  & 0
\\ 
Change in cash & $+\dot{H}$ &  &  &  &  & $-\dot{H}$ & 0 \\ 
Change in bills & $+\dot{\Theta_h}$ &  &  & $+\dot{\Theta_b}$ & $-\dot{\Theta}$ & $+\dot{\Theta_{cb}}$ & 0 \\ 
Change in demand deposits & $+\dot{M_h}$ & \multicolumn{2}{|c|}{$+\dot{M_f}$}
& $-\dot{M}$ &  &  & 0 \\ 
Change in time deposits & $+\dot{D}$ &  &  & $-\dot{D}$ &  &  & 0 \\ 
Change in reserves &  &  &  & $+\dot{R}$ &  & $-\dot{R}$ & 0 \\ \hline
Column sum & $S_h$ & \multicolumn{2}{|c|}{$S_f$} & $S_b$ & $S_g$ & 0 & $p(I-\delta K)$ \\ \hline
Change in net worth & $S_h$ & \multicolumn{2}{|c|}{$S_f+\dot{p}K$} & $S_b$ & 
$S_g$ & 0 & $\dot pK +p\dot K$ \\ \hline
\end{tabular}} \label{table}
\end{table}
%

\subsection{Balance Sheets}

Households distribute their wealth into cash, treasury bills, and demand
deposits. The total amount of cash in circulation is denoted by $H$ and is a
liability for the Central Bank. Treasury bills are short term liabilities of
the government sector and pay an interest rate $r_\theta$, which plays the
role of the main policy rate in the model. For the purposes of this model
they can be thought of as being instantaneously issued or redeemed by the
government to finance its fiscal policy. Because of this feature, their
unit value is deemed to be constant. The total amount of treasuries issued
by the government is denoted by $\Theta$ and is divided into the holdings $%
\Theta_h$ of households, $\Theta_b$ of banks and $\Theta_{cb}$ of the central
bank as specified shortly. 
Demand deposits are liabilities of the banking sector redeemable by cash and
paying an interest rate $r_m$. Observe that we assume for simplicity that
households do not make loans from banks. A more complete model can include
consumer credit in addition to the credit for firms treated in this paper,
but we defer this to further work.

The firm sector produces a homogeneous good used both for consumption and
investment. It utilizes capital with monetary value denoted by $pK$ where $p$ is the
unit price of the homogenous good. The capital stock of firms is partially
financed by loans with total value $L$ at an interest rate $r$. 

The balance sheet of banks consist of demand deposits $M$ and time deposits $%
D$ as liabilities and firm loans $L$, treasury bills $\Theta_b$ and central
bank reserves $R$ as assets. The key feature of fractional reserve banking
is that banks are required to maintain a reserve account with the central bank at the level 
\begin{equation}  \label{required}
R = fM,
\end{equation}
for a constant $0\leq f<1$. 
We assume that banks maintain this required level of reserves by selling and
buying treasury bills to and from the central bank. Observe that no reserve
requirement is assumed for time deposits.

Finally, the public sector is divided into a government that issues bills 
to finance its fiscal deficit (essentially the difference between spending
and taxation) and a central bank that issues cash and reserves as
liabilities and purchases bills as part of its monetary policy.

The column sums along the balance sheet matrix indicate the net worth of
each sector, whereas the row sums are all equal to zero with the exception
of the capital stock, as each financial asset for one sector correspond to a
liability of another. Observe that we assume that the central bank has
constant zero net worth, which in particular implies that it transfers all
profits back to the government.

\subsection{Transactions and Flow of Funds}

Having defined the balance sheet items, the transactions in Table \ref{table}
are self-explanatory and lead to the financial balances, or savings,
indicated as the column sum for each sector. We now describe how these
financial balances are redistributed among the corresponding balance sheet
items for each sector. Starting with the government sector, we have that 
\begin{equation}
S_g = -pG+pT-r_\theta(\Theta_h+\Theta_b),
\end{equation}
showing that government savings are the negative of deficit spending and
interest paid on bills held by the private sector. As this is entirely
financed by net issuance of new bills we have 
\begin{equation}  \label{gov_bills}
\dot\Theta = pG-pT+r_\theta(\Theta_h+\Theta_b) .
\end{equation}
Moving to firms, once depreciation is taken into account, we find from Table %
\ref{table} that savings for firms, after paying wages, taxes, interest on
debt, and depreciation (i.e consumption of fixed capital), are given by 
\begin{equation}
S_f = pY - W -pT-rL +r_mM_f- p\delta K
\end{equation}
and corresponds to the internal funds available for investment. In this
model, the only source for external financing are loans from the banking
sector, so that we have 
\begin{equation}  \label{debt_franke}
\dot L -\dot M_f= p(I-M K) -S_f = pI- \Pi_p,
\end{equation}
where 
\begin{equation}  \label{profits}
\Pi_p=pY-W-pT-rL+r_mM_f
\end{equation}
denotes the after-tax, pre-depreciation profits of the firm sector. The
exact distribution of the difference $(pI- \Pi_p)$ into net new loans and
new deposits depends on portfolio decisions by firms, including a desired
rate of repayment of existing debt. For simplicity, we adopt the
specification in equations (53)-(54) of \cite{GrasselliNguyenHuu2015} with
the repayment rate set to zero, namely, we assume that 
\begin{align}  \label{firm1}
\dot L & = pI +rL \\
\dot M_f &= pY-W+r_mM_f = \Pi_p +rL  \label{firm2}
\end{align}

The flow of funds for households is slightly more involved, as it requires a
choice among different assets. As we can see from Table \ref{table}, the
savings of households are given by 
\begin{equation}
\dot{X}_{h}=S_{h}=pY_{h}-pC,
\end{equation}%
where 
\begin{equation}
pY_{h}=W+r_{m}M_{h}+r_{d}D+r_{\theta }\Theta _{h}+\Delta _{b}
\label{disposable}
\end{equation}%
is the nominal disposable income of households. These savings are then
redistributed among the different balance sheet items held by households so
that 
\begin{equation}
S_{h}=\dot{H}+\dot{\theta}_{h}+\dot{M}_{h}+\dot{D}.
\end{equation}%
To obtain the proportions of savings invested in each type of assets we use
the following modified version of the portfolio equations proposed in
Chapter 10 of \cite{GodleyLavoie2007}: 
\begin{align}
H& =\lambda _{0}X_{h}  \label{tobin1} \\
\Theta _{h}& =(\lambda _{10}+\lambda _{11}r_{\theta }+\lambda
_{12}r_{m}+\lambda _{13}r_{d})X_{h} \\
M_{h}& =(\lambda _{20}+\lambda _{21}r_{\theta }+\lambda _{22}r_{m}+\lambda
_{23}r_{d})X_{h} \\
D& =(\lambda _{30}+\lambda _{31}r_{\theta }+\lambda _{32}r_{m}+\lambda
_{33}r_{d})X_{h}  \label{tobin3}
\end{align}%
subject to the constraints 
\begin{align}
\lambda _{0}+\lambda _{10}+\lambda _{20}+\lambda _{30}& =1 \\
\lambda _{11}+\lambda _{21}+\lambda _{31}& =0 \\
\lambda _{12}+\lambda _{22}+\lambda _{32}& =0 \\
\lambda _{13}+\lambda _{23}+\lambda _{33}& =0
\end{align}%
and the symmetry conditions $\lambda _{ij}=\lambda _{ji}$ for all $i,j$.
These correspond to Tobin's prescription for macroeconomic portfolio
selection, whereby the proportion of the total wealth invested in each class
of assets depends on the rates of returns, with increased demand for one
asset leading to decreased demand for all others.

For our purposes, the important consequence of \eqref{tobin1}--\eqref{tobin3}
is that 
\begin{align}
\dot H &= \lambda_0 \dot X_h = \lambda_0(pY_h - pC)  \label{hh1} \\
\dot\Theta_h &= \lambda_1 \dot X_h = \lambda_1(pY_h - pC)  \label{hh2} \\
\dot M_h &= \lambda_2 \dot X_h = \lambda_2(pY_h - pC)  \label{hh3} \\
\dot D &= \lambda_3 \dot X_h = \lambda_3(pY_h - pC)  \label{hh4}
\end{align}
where 
\begin{align}
\lambda_0 &= 1-(\lambda_{10}+\lambda_{20} +\lambda_{30}) \\
\lambda_1 &=
\lambda_{10}+\lambda_{11}r_\theta+\lambda_{12}r_m-(\lambda_{11}+%
\lambda_{12})r_d \\
\lambda_2 &=
\lambda_{20}+\lambda_{12}r_\theta+\lambda_{22}r_m-(\lambda_{12}+%
\lambda_{22})r_d \\
\lambda_3 &=
\lambda_{30}-(\lambda_{11}+\lambda_{12})r_\theta-(\lambda_{12}+%
\lambda_{22})r_m+(\lambda_{11}+2\lambda_{12}+\lambda_{22})r_d.
\end{align}

The allocations in the remaining assets is now jointly determined by the
interaction between the banking sector and the central bank. To being with,
the reserve requirement \eqref{required} imposes that 
\begin{equation}  \label{dotr}
\dot R = f\dot M = f[\Pi_p+rL+\lambda_2(pY_h - pC)],
\end{equation}
where we used \eqref{firm2} and \eqref{hh3}. Next, the fact that the central
bank transfers all profits to the government sector implies that 
\begin{equation}  \label{cb}
\dot \Theta_{cb} = \dot H + \dot R = (\lambda_0+ f\lambda_2)(pY_h -
pC)+f(\Pi_p+rL),
\end{equation}
where we used \eqref{firm2}, \eqref{hh1} and \eqref{hh3}.

Finally, denoting bank profits by 
\begin{equation}
\Pi_b= rL - r_mM -r_dD+ r_\theta \Theta_b = S_b+\Delta_b
\end{equation}
we see that the the holding of treasury bills by banks satisfies 
\begin{align}
\dot \Theta_b &= S_b +\dot M+\dot D-\dot R-\dot L= \Pi_b-\Delta_b+(1-f)\dot
M +\dot D-\dot L  \label{Theta_dot} \\
& =\Pi_b-\Delta_b+(1-f)[\Pi_p+rL+\lambda_2(pY_h - pC)] +\lambda_3(pY_h - pC)
-pI-rL,  \nonumber
\end{align}
where we used \eqref{firm1}, \eqref{hh3}, and \eqref{dotr}.

At this point it is instructive to observe that it follows from \eqref{hh2}, %
\eqref{cb} and \eqref{Theta_dot} that 
\begin{align*}
\dot \Theta =& \,\, \dot \Theta_h + \dot \Theta_{cb} + \dot \Theta_b =
\lambda_1(pY_h - pC) + (\lambda_0+ f\lambda_2)(pY_h - pC)+f(\Pi_p+rL) \\
& \qquad + \Pi_b-\Delta_b+(1-f)[\Pi_p+rL+\lambda_2(pY_h -
pC)]+\lambda_3(pY_h - pC)-pI -rL \\
=& \,\, (\lambda_0+\lambda_1+\lambda_2+\lambda_3)(pY_h - pC) +
\Pi_p+rL-r_mM-r_dD+r_\theta\Theta_b -\Delta_b-pI \\
=& \,\,\big(W+r_mM_h+r_dD+r_\theta \Theta_h +\Delta_b-pC\big) \\
& +\big(pY-W-pT-rL+r_mM_f\big)+rL-r_mM-r_dD+r_\theta\Theta_b -\Delta_b-pI \\
=& \,\, pG-pT+r_g(\Theta_h+\Theta_b) ,
\end{align*}
in accordance with \eqref{gov_bills}.

\subsection{Additional Behavioural Assumptions}

To complete the model, we need to specify several additional behavioural
assumptions for each sector. For this, let us first introduce the following
intensive variables 
\begin{align}
\omega &= \frac{W}{pY}, \quad h = \frac{H}{pY}, \quad \theta_h = \frac{%
\Theta_h}{pY}, \quad d = \frac{D}{pY} \\
\quad m_h &= \frac{M_h}{pY}, \quad m_f = \frac{M_f}{pY}, \quad \ell = \frac{L%
}{pY}, \quad \theta_b = \frac{\Theta_b}{pY}.
\end{align}
In addition, let the total working age population be denoted by $N$ and the
number of employed workers by $E$. We then define the productivity per
worker $a$, the employment rate $e$ and the nominal wage rate as 
\begin{equation}  \label{eq:productivity}
a =\frac{Y}{E}, \qquad e = \frac{E}{N} = \frac{Y}{aN}, \qquad \mathrm{w}=%
\frac{W}{E},
\end{equation}
whereas the unit cost of production, defined as the wage bill divided by
quantity produced, is given by 
\begin{equation}  \label{eq:unit_cost}
u_c=\frac{W}{Y}=\frac{\mathrm{w}}{a}.
\end{equation}
We will assume throughout that productivity and workforce grow exogenously
according to the dynamics 
\begin{equation}  \label{eq:productivity and population}
\frac{\dot a}{a} = \alpha, \qquad \frac{\dot N}{N} = \beta.
\end{equation}

\vspace{0.2in} \noindent \textbf{Wage-price dynamics:} For the price
dynamics we assume that the long-run equilibrium price is given by a
constant markup $m\ge 1$ times unit labor cost, whereas observed prices
converge to this through a lagged adjustment with speed $\eta_p>0$. 
Using the fact that the instantaneous unit labour cost is given by $%
u_c=\omega p$, we obtain: 
\begin{equation}  \label{inflation}
\frac{\dot p}{p}=\eta_p\left(m\frac{u_c}{p}-1\right)=\eta_p(m\omega-1):=i(%
\omega)
\end{equation}

We assume that the wage rate $\mathrm{w}$ follows the dynamics 
\begin{equation}  \label{eq:wage 3}
\frac{\dot{\mathrm{w}}}{\mathrm{w}}=\Phi(e)+\gamma\frac{\dot p}{p}
\end{equation}
for a constant $0\leq\gamma\leq 1$. This assumption states that workers
bargain for wages based on the current state of the labour market through
the Philips curve $\Phi$, but also take into account the observed inflation
rates. The constant $\gamma$ represents the degree of money illusion, with $%
\gamma=1$ corresponding to the case where workers fully incorporate
inflation in their bargaining. For the Philips curve, we assume that $%
\Phi(e)\to+\infty$ as $e\to 1$ in order to prevent the employment rate from
going above unit.

\vspace{0.2in} \noindent \textbf{Fiscal policy:} We consider the simplest
case of real government spending and taxation given by 
\begin{align}
G & = g Y  \label{spend} \\
T &= t Y  \label{tax}
\end{align}
for constants $g$ and $t$.

\vspace{0.2in} \noindent \textbf{Investment, production and consumption:} As
in the \cite{Keen1995} model, we assume that the relationship between capital and
output is given by $Y=\frac{K}{\nu }$ for a constant capital-to-output
ratio. There are many ways to relax this condition, for example by
introducing a variable utilization rate as in \cite{GrasselliNguyen2018},
but we will not pursue them here. Capital itself is assumed to change
according to 
\begin{equation}
\dot{K}=I-\delta K,  \label{eq:capital}
\end{equation}%
where $\delta $ is a depreciation rate. Moreover, we assume that real
investment is given by 
\begin{equation}
I=\kappa (\pi )Y\;,  \label{eq:investment}
\end{equation}%
for a function $\kappa $ of the profit share 
\begin{equation}
\pi =\frac{\Pi _{p}}{pY}=1-t-\omega -r\ell +r_{m}m_{f}.
\label{eq:franke expected profit}
\end{equation}%
Using \eqref{eq:investment}, \eqref{eq:capital} and $Y=\frac{K}{\nu }$, we
find that the growth rate of real output is given by 
\begin{equation}
g_{Y}(\pi ):=\frac{\dot{Y}}{Y}=\frac{\dot{Y}}{Y}=\frac{\kappa (\pi )}{\nu }%
-\delta .  \label{growth_keen}
\end{equation}%
Furthermore, still in line with the original Keen model, we assume that all
output is sold, so that there are no inventories or any difference between
supply and demand. Accordingly, real consumption of households is given by 
\begin{equation}
C=Y-G-I=(1-g-\kappa (\pi ))Y.
\end{equation}


%

\vspace{0.2in} \noindent \textbf{Bank dividends:} There are many alternative
definitions of bank behaviour that are compatible with the accounting
structure described in Table \ref{table}. For example, in \cite%
{GodleyLavoie2007}, it is assume throughout the book that all bank profits
are immediately distributed to households, so that the financial balances of
banks is always identically zero and, consequently, the net worth of banks
is kept constant. The problem with this approach is that, in a growing
economy, it leads to vanishing capital ratios, as loans and deposits
continue to grow while the equity of the bank remains constant. In \cite%
{GrasselliLipton2018} this was remedied by assuming that banks target a
regulatory capital ratio $k_r$ and distribute profits accordingly. For the
present model, this assumption translates into 
\begin{equation}  \label{capital_r}
X_b = k_r (\rho_L L + \rho_g \Theta_b + \rho_r R),
\end{equation}
that is, we assume that banks distribute enough dividends to keep equity
equal to a multiple $k_r$ of risk-weighted assets. For simplicity, we take $%
\rho_L = 1$ and $\rho_g = \rho_r =0$, but the same general argument applies
to arbitrary risk weights. Because bank savings need to equal the change in
bank equity (i.e net worth), we have that 
\begin{equation}  \label{bank_saving}
S_b = \dot X_b = k_r \dot L = k_r (pI+rL),
\end{equation}
which in turn implies that bank dividends are 
\begin{equation}  \label{dividends}
\Delta_b = \Pi_b- k_r (pI+rL).
\end{equation}
Looking back at the expressions involving bank dividends, we see from %
\eqref{disposable} that nominal disposable income for households is equal to 
\begin{equation}  \label{disposable2}
pY_h= W+(1-k_r)rL -r_mM_f+ r_\theta(\Theta_h+ \Theta_b) - k_r pI
\end{equation}
and from \eqref{Theta_dot} that the holding of bills by banks satisfies 
\begin{equation}  \label{Theta_dot_3}
\dot\Theta_b = (1-f)[\Pi_p+\lambda_2(pY_h - pC)]+\lambda_3(pY_h -
pC)-(1-k_r)pI+(k_r-f)rL.
\end{equation}



\subsection{The main dynamical system}


The dynamics for the wage share $\omega=\mathrm{{w}/(pa)}$ obtained from %
\eqref{eq:productivity and population}, \eqref{inflation} and 
\eqref{eq:wage
3} is 
\begin{equation}  \label{wage_keen_modified}
\frac{\dot\omega}{\omega}= \frac{\dot{\mathrm{w}}}{\mathrm{w}}-\frac{\dot p}{%
p}-\frac{\dot a}{a}=\Phi(e)-\alpha-(1-\gamma)i(\omega),
\end{equation}

For the employment rate $e=Y/(aN)$, we use 
\eqref{eq:productivity and
population}, \eqref{growth_keen} to obtain 
\begin{equation}
\frac{\dot{e}}{e} = \frac{\dot{Y}}{Y} - \frac{\dot a}{a}-\frac{\dot N}{N}= 
\frac{\kappa(\pi)}{\nu}-\delta -\alpha-\beta \;.  \label{employment_dynamics}
\end{equation}

For the household variables $h=H/(pY)$, $\theta _{h}=\Theta _{h}/(pY)$, $%
m_{h}=M_{h}/(pY)$ and $d=D/(pY)$, we use \eqref{hh1}-\eqref{hh4} to obtain 
\begin{align}
\frac{\dot{h}}{h}=& \frac{\dot{H}}{H}-\frac{\dot{p}}{p}-\frac{\dot{Y}}{Y}=%
\frac{\lambda _{0}(pY_{h}-pC)}{H}-\Gamma (\omega ,\ell ,m_{f})
\label{high_keen} \\
\frac{\dot{\theta}_{h}}{\theta _{h}}=& \frac{\dot{\Theta}_{h}}{\Theta _{h}}-%
\frac{\dot{p}}{p}-\frac{\dot{Y}}{Y}=\frac{\lambda _{1}(pY_{h}-pC)}{\Theta
_{h}}-\Gamma (\omega ,\ell ,m_{f})  \label{bills_h_keen} \\
\frac{\dot{m}_{h}}{m_{h}}=& \frac{\dot{M}_{h}}{M_{h}}-\frac{\dot{p}}{p}-%
\frac{\dot{Y}}{Y}=\frac{\lambda _{2}(pY_{h}-pC)}{M_{h}}-\Gamma (\omega ,\ell
,m_{f})  \label{deposit_keen} \\
\frac{\dot{d}}{d}=& \frac{\dot{D}}{D}-\frac{\dot{p}}{p}-\frac{\dot{Y}}{Y}=%
\frac{\lambda _{3}(pY_{h}-pC)}{D}-\Gamma (\omega ,\ell ,m_{f})
\label{deposit_keen}
\end{align}%
where%
\begin{equation}
\Gamma (\omega ,\ell ,m_{f})=g_{Y}(\pi )+i(\omega )=\frac{\kappa (\pi )}{\nu 
}-\delta +i(\omega )
\end{equation}%
Similarly, for the firm variables $\ell =L/(pY)$ and $m_{f}=M_{f}/(pY)$, we
use \eqref{firm1}-\eqref{firm2} to obtain 
\begin{align}
\frac{\dot{\ell}}{\ell }& =\frac{\dot{L}}{L}-\frac{\dot{p}}{p}-\frac{\dot{Y}%
}{Y}=\frac{pI}{L}+r-\Gamma (\omega ,\ell ,m_{f})  \label{debt_keen} \\
\frac{\dot{m}_{f}}{m_{f}}& =\frac{\dot{M}_{f}}{M_{f}}-\frac{\dot{p}}{p}-%
\frac{\dot{Y}}{Y}=\frac{\Pi _{p}}{M_{f}}+\frac{r\ell }{m_{f}}-\Gamma (\omega
,\ell ,m_{f})  \label{deposit_keen_f}
\end{align}%
Finally, for the ratio of bank holdings of bills $\theta _{b}=\Theta
_{b}/(pY)$, we can use \eqref{Theta_dot_3} to obtain 
\begin{align}
\frac{\dot{\theta}_{b}}{\theta _{b}}& =\frac{\dot{\Theta}_{b}}{\Theta _{b}}-%
\frac{\dot{p}}{p}-\frac{\dot{Y}}{Y} =\frac{(k_{r}-f)r\ell }{\theta _{b}} -\Gamma (\omega,\ell ,m_{f})\label{bills_b_keen} \\
& \qquad +\frac{(1-f)\Pi _{p}+[\lambda
_{3}+(1-f)\lambda _{2}](pY_{h}-pC)-(1-k_{r})pI}{\Theta _{b}}  \nonumber
\end{align}

We then find that \eqref{wage_keen_modified}--\eqref{employment_dynamics}
and \eqref{debt_keen}--\eqref{deposit_keen_f} lead to the following system
of ordinary differential equations: 
\begin{equation}
\left\{ 
\begin{array}{ll}
\dot{\omega} & =\left[ \Phi (e)-(1-\gamma )i(\omega )-\alpha \right] \omega 
\\ 
\dot{e} & =\left[ \frac{\kappa (\pi )}{\nu }-\alpha -\beta -\delta \right] e
\\ 
\dot{\ell} & =\left[ r-\Gamma (\omega ,\ell ,m_{f})\right] \ell +\kappa (\pi
) \\ 
\dot{m}_{f} & =\left[ r_{m}-\Gamma (\omega ,\ell ,m_{f})\right] m_{f}-\omega
+1-t%
\end{array}%
\right.   \label{keen}
\end{equation}%
where 
\begin{align}
\pi & =1-t-\omega -r\ell +r_{m}m_{f}  \label{profit_cf} \\
i(\omega )& =\eta _{p}(m\omega -1)  \label{inflation_cf}
\end{align}

To solve \eqref{keen}, it is necessary to specify the behavioural functions $%
\Phi (\cdot )$ and $\kappa (\cdot )$. For the Philips curve we follow \cite%
{GrasselliNguyenHuu2015} and choose 
\begin{equation}
\Phi (e)=\frac{\phi _{1}}{(1-e)^{2}}-\phi _{0}
\end{equation}%
for constants $\phi _{0},\phi _{1}$ specified in Table \ref{parameters_table}%
. For the investment function, we follow the more recent work of \cite%
{NguyenHuuPottier2017} and use 
\begin{equation}
\kappa (\pi )=\kappa _{0}+\frac{\kappa _{1}}{(\kappa _{2}+\kappa
_{3}e^{-\kappa _{4}\pi })^{\xi }}
\end{equation}%
that is to say, a generalized logistic function with parameters given in
Table \ref{parameters_table}.

Once the main system \eqref{keen} is solved for the state variables $(\omega
,e,\ell ,m_{f})$, we can use them to solve the following auxiliary system for
the variables $(\theta _{h},\theta _{b})$ derived from \eqref{bills_h_keen}
and \eqref{bills_b_keen}: 
\begin{equation}
\left\{ 
\begin{array}{ll}
\dot{\theta}_{h} & =-\Gamma (\omega ,\ell ,m_{f})\theta _{h}+\lambda _{1}\Xi
(\omega ,\ell ,m_{f},\theta _{p}) \\ 
\label{private}\dot{\theta}_{b} & =-\Gamma (\omega ,\ell ,m_{f})\theta
_{b}+\left( \lambda _{3}+(1-f)\lambda _{2}\right) \Xi (\omega ,\ell
,m_{f},\theta _{p}) \\ 
& \qquad -(1-k_{r})\kappa (\pi )+(1-f)\pi +(k_{r}-f)r\ell 
\end{array}%
\right. 
\end{equation}%
where $\theta _{p}$ is total private holding of government bills: 
\begin{equation}
\theta _{p}=\theta _{h}+\theta _{b},
\end{equation}%
and%
\begin{equation}
\Xi (\omega ,\ell ,m_{f},\theta _{p})=g-t-\pi +r_{\theta }(\theta
_{h}+\theta _{b})+(1-k_{r})\kappa (\pi )-k_{r}r\ell 
\end{equation}%
we can then find the following remaining variables separately by solving
each of the following auxiliary equations: 
\begin{align}
\dot{h}& =-\Gamma (\omega ,\ell ,m_{f})h+\lambda _{0}\Xi (\omega ,\ell
,m_{f},\theta _{p}) \\
\dot{m}_{h}& =-\Gamma (\omega ,\ell ,m_{f})m_{h}+\lambda _{2}\Xi (\omega
,\ell ,m_{f},\theta _{p}) \\
\dot{d}& =-\Gamma (\omega ,\ell ,m_{f})d+\lambda _{3}\Xi (\omega ,\ell
,m_{f},\theta _{p})
\end{align}

The system \eqref{keen} is very similar to the system analyzed in Section 4 of 
\cite{GrasselliNguyenHuu2015} if one sets the speculative flow $F=0$ in
their equation (46). We therefore do not repeat the analysis of the
equilibrium points of \eqref{keen}, except for observing that it admits an
interior equilibrium characterized by a profit share defined as 
\begin{equation}  \label{pi_equi}
\overline\pi = \kappa^{-1}(\nu(\alpha+\beta+\delta))
\end{equation}
and corresponding to non-vanishing wage share and employment rate, finite
private debt, and a real growth rate of 
\begin{equation}  \label{g_equi}
\kappa(\overline\pi) = \alpha+\beta.
\end{equation}
In addition, system \eqref{keen} admits a variety of equilibria
characterized by infinite debt ratios and a real growth rate converging to $%
\kappa_0/\nu-\delta<0$. In Section \ref{numerical} we explore the properties
of these different equilibria in the context of the present model.

\section{Narrow banking}

\label{narrow}

Several alternative definitions for narrow banking have been summarized in 
\cite{Pennacchi2012}, ranging from the familiar Full Reserve Banking
advocated in the Chicago Plan to much less recognizable forms of `banking',
such as Prime Money Market Mutual Funds (PMMMF). Common to all the
definitions is a separation between loans and demand deposits.
In what follows, we focus on a specific example of
such separation, namely by dividing the banking sector into Full Reserve
Banking and Lending Facilities.

\subsection{Full Reserve Banking}

The simplest form of narrow banking corresponds to financial institutions
that have only demand deposits as liabilities and are required to hold an
equal amount of reserves as assets, which in the context of the model of
Section \ref{fractional} this corresponds to setting $f=1$. These institutions
can, in principle, also hold cash or excess reserves as assets in addition
to required reserves, with the difference between total assets and demand
deposits corresponding to shareholder equity, or net worth in the notation
of the previous section. For simplicity, in accordance with assigning a risk
weight $\rho_r=0$ to reserves, we assume that Full Reserve Banks maintain
zero net worth, so that required reserves equal demand deposits at all times.%
\footnote{%
Observe that the assumption of zero net worth is made throughout in \cite%
{GodleyLavoie2007} for the entire banking sector, not only for narrow banks.}
In practice, even a bank holding only reserves as assets should have a small
positive net worth to absorb losses due to operational risk, but we will
neglect this effect here. Similarly, because we are assuming zero-interest
on reserves, in practice a full reserve bank would need to charge a service
fee in order to be able to pay interest on demand deposits and generate a
profit. We neglect this effect also and assume that $r_m=0$ so that our full
reserve bank operates with zero profit.\footnote{%
Notice that we have been ignoring operational costs of the banking sector
all along, for example by assuming that they pay no wages to employees, and
the assumption of zero profits for a full reserve bank is not much stronger.}

In other words, a Full Reserve Bank in our model corresponds to the
following balance sheet structure: 
\begin{align*}
\mbox{Assets: } & R \\
\mbox{Liabilities: } & M_h + M_f \\
\mbox{Net Worth: } & X^{1}_b = R - (M_h+M_f) = 0
\end{align*}

\subsection{Lending Facilities}

\label{lending}

These correspond to a financial institution that holds treasury bills and
loans as assets and time deposits as liabilities. At an operational level,
in the context of the model of Section \ref{fractional} a lending facility
acquires time deposits when a household decides to reallocate part of its
wealth away from other assets. For example, a household can transfer funds
from its demand deposit account with a Full Reserve Bank into a time deposit
account in a Lending Facility. This is accompanied by a transfer of reserves
from the Full Reserve Bank to the Lending Facility. Similarly operations
take place when households reallocate their wealth from cash and government
bills into time deposits, all leading to an increase in reserves temporarily
held by the Lending Facility.

These excess reserves (as there is no reserve requirements for time
deposits) can then either be used to to purchase bills from the central bank
or to create a new loan, which results in the excess reserves being
transferred back to the Full Reserve Bank, but this time as a demand deposit
for the borrowing firm.

The key feature of narrow banking is that the Lending Facility is not able
to create new time deposits simply by creating new loans. Instead, the
Lending Facility needs to first obtain excess reserves in the amount of the
new loan. One way to obtain excess reserves is by attracting time deposits
as described above. Another consists of selling government bills to the
central bank. In other words, lending corresponds to an asset swap, without
any expansion of the balance sheet of the Lending Facility.

In addition, the Lending Facility is constrained by the minimal capital
requirement \eqref{capital_r}, which in this case reduces to 
\begin{equation}  \label{theta_bound}
\Theta_b = D-(1-k_r)L.
\end{equation}
We can see that whenever $D$ drops below $(1-k_r)L$, the only way for the
Lending Facility to meet its minimal capital requirement is by borrowing
from the government.\footnote{%
Observe that this is essentially the same mechanism explained in the section
``Bank lending under the Sovereign Money system" of \cite{KPMG2016}.}

To summarize, a Lending Facility in our model corresponds to the following
balance sheet structure: 
\begin{align*}
\mbox{Assets: } & L+\Theta_b \\
\mbox{Liabilities: } & D \\
\mbox{Net Worth: } & X^{2}_b = L+\Theta_b-D = k_r L
\end{align*}

As we mentioned in Section \ref{introduction}, a Full Reserve Bank and a Lending Facility can be owned and managed 
as a single bank with two distinct business lines. The key point is that demand deposits need to be matched with 
an equal amount of central bank reserves, regardless of the remaining mix of assets and liabilities of the bank. 

For our purposes, as narrow banking regime is therefore characterized by a banking sector with for which $f=1$ in \eqref{required}. In the 
next section we compare the properties of fractional and full reserve banking through a series of numerical examples.

%
%

\section{Numerical Experiments}

\label{numerical}

We perform four experiments to demonstrate the properties of the model under
different reserve requirements. In all cases we use the base parameters
shown in Table \ref{parameters_table}. Details on the parameters used for
the wage, employment, and inflation parts of the model, namely $\alpha,
\beta, \eta_p, m, \nu$ and $\delta$ can be found in \cite%
{GrasselliMaheshwari2018}, whereas an in-depth discussion of the properties
of the investment function and its parameters, namely $\kappa_i$, $%
i=1,\ldots,4$ and $\xi$ can be found in \cite{NguyenHuuPottier2017}. The
remaining parameters, namely the interest rates $r, r_D, r_\theta$ and $r_m$%
, the capital adequacy ratio $k_r$, and the constants $g$ and $t$ related to
government spending and taxation are used for illustration only and are
based on recent representative values in advanced economies. Initial
conditions for each experiment are indicated in the figures showing the
results.

\begin{figure}[!ht]
\centering
\includegraphics[width=\textwidth]{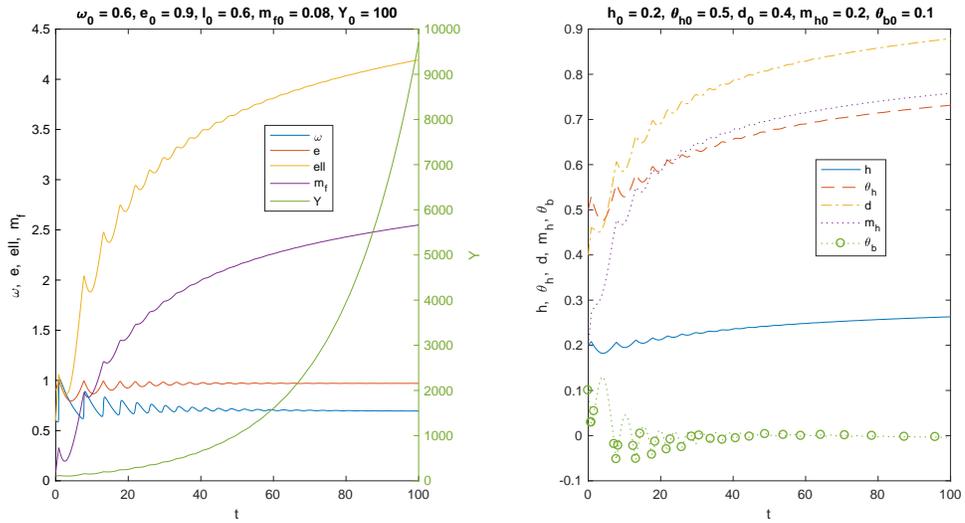}
\caption{Solution of the model \eqref{keen} with fractional reserve ratio $%
f=0.1$ and remaining parameters as in Table \protect\ref{parameters_table}.
With a moderate value for the initial loan ratio $\ell_0=0.6$ we observe
convergence to an interior equilibrium.}
\label{fractional_finite_fig}
\end{figure}

\begin{example} {\bf (fractional reserve banking with finite debt)}

\label{fractional_finite}

\vspace{0.1in}
\noindent
We begin with an example of fractional reserve banking where the main system \eqref{keen} reaches an interior equilibrium. We take $f=0.1$ as the required reserve ratio and choose a moderate level of loan ratio 
$\ell_0=0.6$ as an initial condition. As shown in the left panel of Figure \ref{fractional_finite_fig}, the state variables for \eqref{keen} converge to the equilibrium 
\begin{equation}
(\hat\omega,\hat\lambda,\hat \ell,\hat m_f)=(0.6948,0.9706,4.1937,0.7577).
\end{equation}
The profit share corresponding to this equilibrium according to \eqref{profit_cf} is 
\[\hat\pi = 0.1249,\] 
leading to a growth rate of real output of 
$g(\hat\pi)=0.0451$ according to \eqref{growth_keen}. These are very good approximations to the theoretical values 
$\overline\pi = 0.1248$ and $g(\overline\pi)=0.0450$ obtained from \eqref{pi_equi} and \eqref{g_equi}. 

The right panel of Figure \ref{fractional_finite_fig} shows convergence of the other variables of the model to finite values. In particular, observe that although the loan ratio $\ell$ and the deposit ratios $d, m_f$ and $m_h$ all increase before stabilizing at their equilibrium values, the ratio $\theta_b$ of bills held by the banking sector remains close to its small initial value $\theta_b = 0.1$, indicating the usual money market interactions between banks and the central bank.  

\end{example}

\begin{figure}[!ht]
\centering
\includegraphics[width=\textwidth]{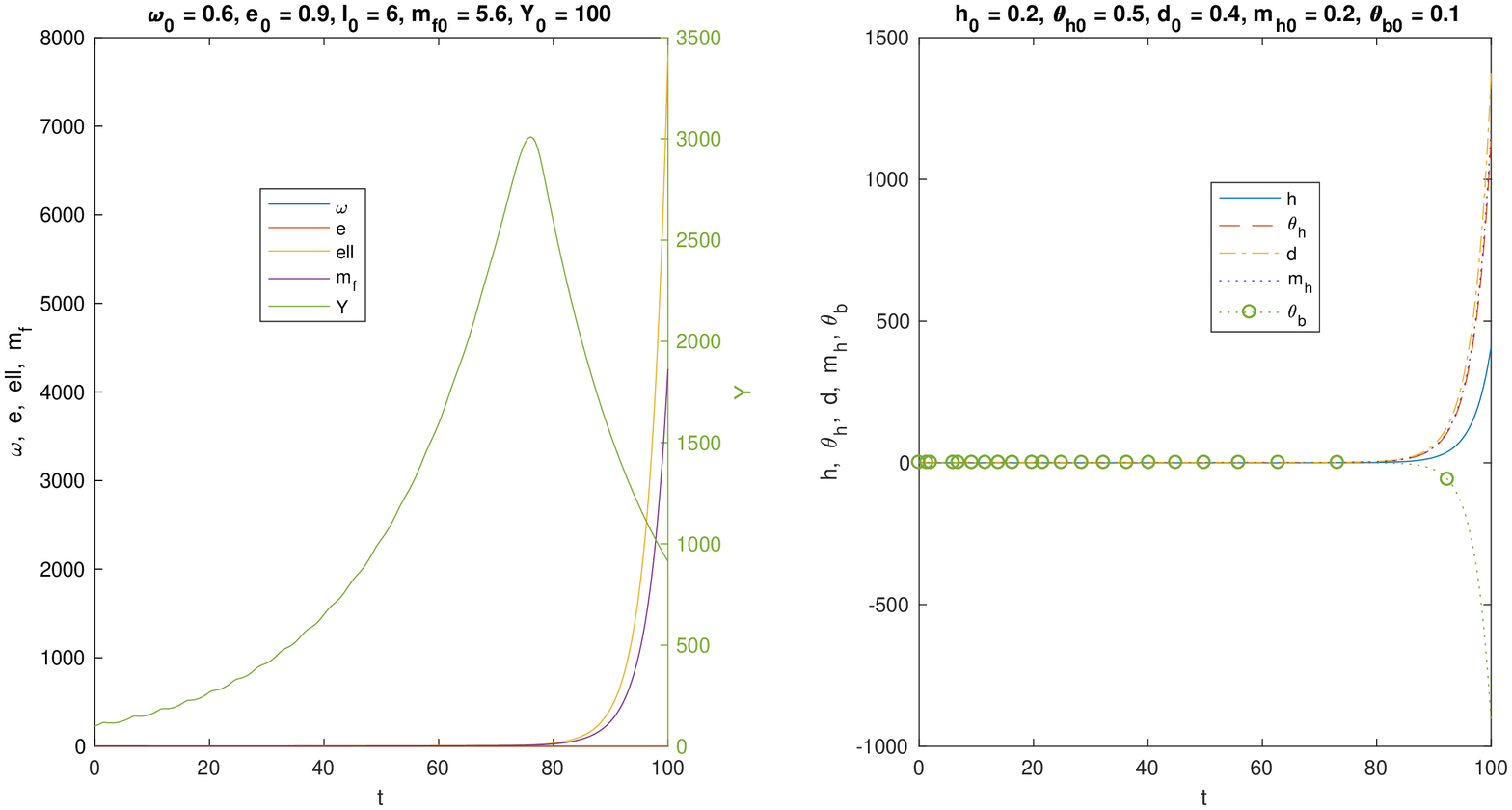}
\caption{Solution of the model \eqref{keen} with fractional reserve ratio $%
f=0.1$ and remaining parameters as in Table \protect\ref{parameters_table}.
With a high value for the initial loan ratio $\ell_0=6$ we observe
convergence to an equilibrium with infinite private debt.}
\label{fractional_infinite_fig}
\end{figure}

\begin{example} {\bf (fractional reserve banking with explosive debt)}

\label{infinite_fractional}

\vspace{0.1in}
\noindent
We consider next an example of fractional reserve banking where the main system \eqref{keen} approaches an equilibrium with infinite private debt and vanishing wage share and employment rate. As before, we take $f=0.1$ as the required reserve ratio but modify the initial loan ratio to $\ell_0=6$, that is to say, ten times larger than in the previous example. Admittedly, this is an extreme initial condition\footnote{The level of domestic credit to the private sector as a proportion of GDP (which is approximated by $\ell$ in our model) was approximately 1.3 for the entire world in 2016 (up from 0.5 in 1960) and only larger than 2 for Cyprus. Source: IMF, International Financial Statistics (https://data.worldbank.org/indicator/FS.AST.PRVT.GD.ZS)}, chosen here for illustrative purposes.  The key point is that, as shown in \cite{GrasselliNguyenHuu2015}, explosive equilibria of this type for \eqref{keen} are locally stable for a wide range of parameters, and therefore cannot be ignored from the outset. 

As shown in the left panels of Figure \ref{fractional_infinite_fig}, both the loan and deposit ratios $\ell$ and $m_f$ for firms eventually explode to infinity in this example, dragging the economy down with 
a growth rate $-0.0522$ and causing the wage share and employment rate to converge to zero. The remaining variables of the model are shown in the right panel of Figure \ref{fractional_infinite_fig}, where we can see the household holdings of cash, demand and time deposits, and bills all exploding to infinity. Characteristically, we see that $\theta_b\to -\infty$, indicating that the banking sector 
needs to {\em borrow} from the government in order to keep its equity ratio at the desired level $k_r$. 

\end{example}

\begin{figure}[!ht]
\centering
\includegraphics[width=\textwidth]{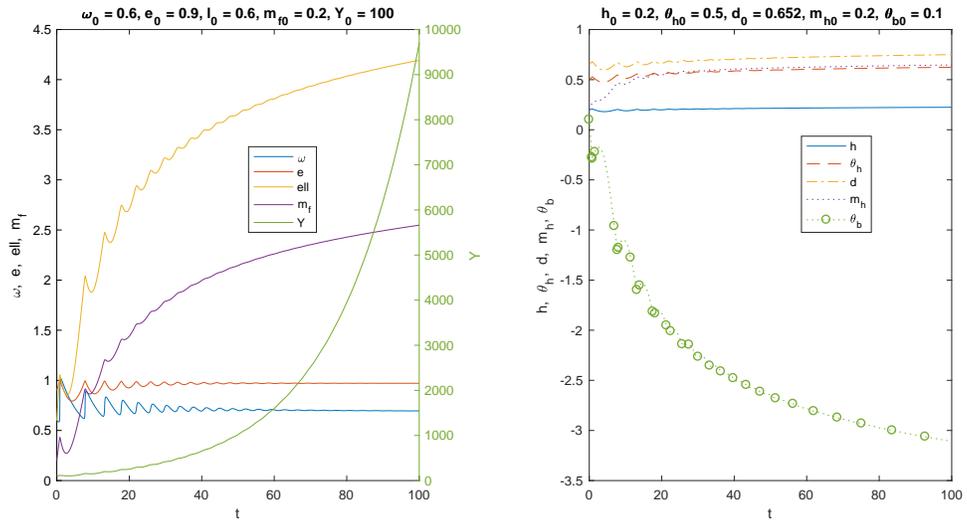}
\caption{Solution of the model \eqref{keen} with full reserve ratio $f=1$
and remaining parameters as in Table \protect\ref{parameters_table}. With a
moderate value for the initial loan ratio $\ell_0=0.6$ we observe
convergence to an interior equilibrium. Observe the negative values for $%
\protect\theta_b$ throughout the period.}
\label{narrow_finite_fig}
\end{figure}

\begin{example} {\bf (narrow banking with finite equilibrium)}

\vspace{0.1in}
\noindent
In this example we use the same parameters as in Example \ref{fractional_finite} with the only difference that $f=1$, namely, we impose a 100\% reserve requirement. We also use the same initial 
conditions as in Example \ref{fractional_finite}, except for $mf_0$ and $d_0$, which need to be calculated differently to achieve the required capital ratio in this case. 

The left panel of Figure \ref{narrow_finite_fig} shows the state variables of \eqref{keen} converging to essentially the same equilibrium as before, namely 
\begin{equation}
(\hat\omega,\hat\lambda,\hat \ell,\hat m_f)=(0.6948,0.9706,4.1929,0.6462),
\end{equation}
corresponding to a profit share and growth rates that are identical up to four decimal places. Notably, narrow banking does {\em not} lead to any loss in equilibrium growth for the economy. 

The only significant departure from the fractional banking case of Example \ref{fractional_finite} is that $\theta_b$ drops to negative values almost immediately and continues to become more and 
more negative as $\ell$ increases towards its equilibrium value. In other words, in the full reserve case the banking sector needs to borrow more from the government in order to increase its lending to the private sector.
\end{example}

\begin{figure}[!ht]
\centering
\includegraphics[width=\textwidth]{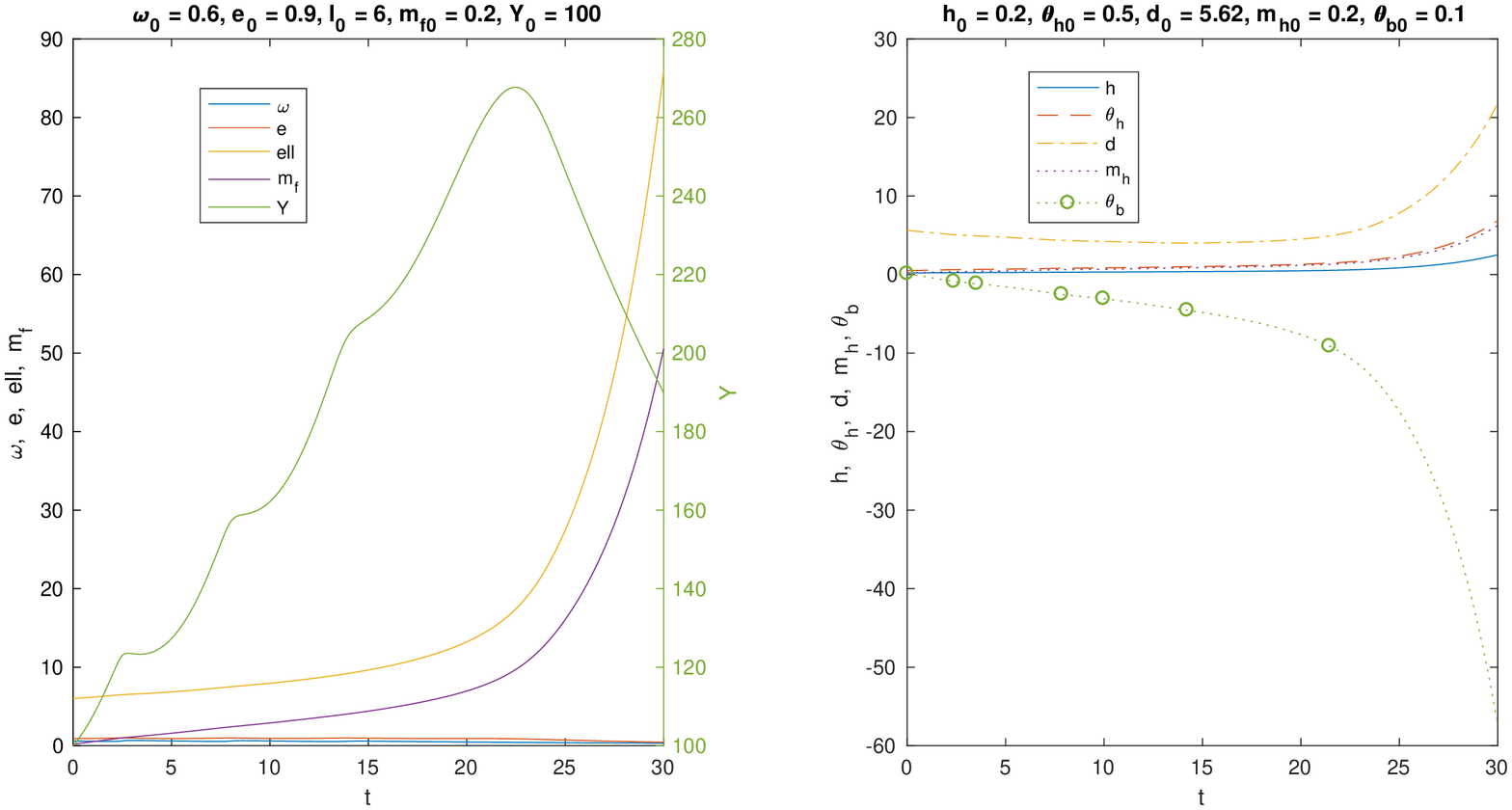}
\caption{Solution of the model \eqref{keen} with full reserve ratio $f=1$
and remaining parameters as in Table \protect\ref{parameters_table}. With a
high value for the initial loan ratio $\ell_0=6$ we observe convergence to
an equilibrium with infinite private debt.}
\label{narrow_infinite_fig}
\end{figure}

\begin{example} {\bf (narrow banking with explosive debt)}

\vspace{0.1in}
\noindent
In this example we use $f=1$ and the same values for parameters and initial conditions as in Example \ref{infinite_fractional} except for $mf_0$ and $d_0$, which again need to be calculated differently to achieve the required capital ratio in the full reserve case. 

As in Example \ref{infinite_fractional}, we see in the left panel of Figure \ref{narrow_infinite_fig} that the high initial level of debt for the firm sector leads to an explosive 
behaviour for the variables $\ell$ and $m_f$ in system \eqref{keen} and corresponding collapse of output, wages and employment. The essential difference is that this occurs much earlier in the 
narrow banking case, namely output begins to decrease shortly after 20 years of debt accumulation, as oppose to after nearly 70 years in the fractional banking case. Moreover, as we can see in 
the right panel of Figure \ref{narrow_infinite_fig}, the reliance of the banking sector on {\em borrowing} from the government is much more pronounced in the narrow banking case, with $\theta_b$ surpassing (i.e becoming more negative than) $-1$ within a few years. 

\end{example}

\section{Conclusions}
\label{conclusion}

In this paper we have considered two stock-flow consistent economic models:
(A) one with the traditional fractional reserve banking sector and (B) one with
the narrow banking sector. We analyzed their similarities and differences
and demonstrated that both can operate in a satisfactory fashion, with a
narrow banking system exhibiting features that allow for better monitoring and 
prevention of crises by regulators. Crucially, the version of the model with a 100\% reserve requirement for 
demand deposits did {\em not} suffer from any loss of economic growth when compared with the fractional 
reserve version. 

Several improvements can be made to the base model presented here, adding realism 
at the expense of tractability, as expected. One relates to the usual criticism that the Keen model 
does not incorporate a realistic consumption function, variable utilization of capital, and inventory management. All 
of these features can be added to the current model essentially in the same way as in \cite{GrasselliNguyen2018}, with 
the corresponding increase in dimensionality for the system. Similarly, adding stochasticity to some of the underlying economic variables, such 
as productivity growth, is an important open task that should be carried out along the lines developed in \cite{NguyenHuuCostaLima2014} and 
\cite{Lipton2016b}. Specifically related to the topic of this paper, a natural extension consists in restricting the supply of credit to firms when 
the level of government {\em lending} to the private sector is deemed too high, as a potential stabilization policy. This can be done with the addition of a
credit rationing mechanism similar to what is proposed in \cite{DafermosNikolaidiGalanis2017}. In a similar vein, default by both firms and banks is a
very important aspect that needs to be incorporated into the model. 

In light of the oversized role that banking and finance play in modern economies, effective regulation of the banking sector remains the number one priority for achieving 
systemic stability. Narrow banking is a compelling policy tool with a long pedigree but poorly understood properties. While advances in technology make the implementation of 
narrow banking more feasible than it has ever been, concerns about the macroeconomic consequences of the policy persist, in particular with respect to growth. For example, voters in a recent referendum in Switzerland resoundingly rejected a narrow banking proposal largely because of the uncertainties surrounding the idea\footnote{See 
https://www.reuters.com/article/us-swiss-vote-sovereign/swiss-voters-reject-campaign-to-radically-alter-banking-system-idUSKBN1J60C0}
We hope to have contributed to the discussion by showing that the advantages of narrow banking merit serious consideration by regulators and policy makers. 

\bibliographystyle{apalike}
\bibliography{finance}

\newpage
\appendix

\section{Parameters for numerical simulations}

\label{parameters_section}

The baseline parameters for our simulations are provided in Table \ref%
{parameters_table}. Alternative values for some specific parameters are
provided in the legend of each figure.

\begin{table}[!ht]
\begin{center}
\begin{tabular}{c|l|l}
Symbol & Value & Description \\ \hline
$r$ & 0.04 & interest rate on loans \\ 
$r_D$ & 0.02 & interest rate on time deposits \\ 
$r_\theta$ & 0.012 & interest rate on bills \\ 
$r_m$ & 0.01 & interest rate on demand deposits \\ 
$\lambda_0$ & 0.1 & proportion of households savings invested in cash \\ 
$\lambda_{i0}$ & 0.3 & portfolio parameters for households ($i=1,2,3$) \\ 
$\lambda_{11}$ & 4 & portfolio parameter for households \\ 
$\lambda_{12}$ & -1 & portfolio parameter for households \\ 
$\lambda_{22}$ & 2 & portfolio parameter for households \\ 
$\alpha$ & 0.025 & productivity growth rate \\ 
$\beta$ & 0.02 & population growth rate \\ 
$\eta_p$ & 0.35 & adjustment speed for prices \\ 
$m$ & 1.6 & markup factor \\ 
$\gamma$ & 0.8 & inflation sensitivity in the bargaining equation \\ 
$g$ & 0.2 & government spending as a proportion of output \\ 
$t$ & 0.08 & taxes as a proportion of output \\ 
$\nu$ & 3 & capital-to-output ratio \\ 
$\delta$ & 0.05 & depreciation rate \\ 
$k_r$ & 0.08 & capital adequacy ratio \\ 
$\phi_0$ & 0.0401 & Philips curve parameter \\ 
$\phi_1$ & $6.41 \times 10^{-5}$ & Philips curve parameter \\ 
$\kappa_0$ & -0.0056 & investment function lower bound \\ 
$\kappa_1$ & 0.8 & investment function upper bound \\ 
$\kappa_2$ & 1 & investment function parameter \\ 
$\kappa_3$ & 2 & investment function parameter \\ 
$\kappa_4$ & 10 & investment function parameter \\ 
\ $\xi$ & 4 & investment function parameter \\ \hline
\end{tabular}
\end{center}
\caption{Baseline parameter values}
\label{parameters_table}
\end{table}

\newpage
\section{Figures}

\end{document}